\documentstyle[epsf]{article}
\newcommand{\bee}{\begin{equation}}
\newcommand{\ee}{\end{equation}}
\newcommand{\beea}{\begin{eqnarray}}
\newcommand{\eea}{\end{eqnarray}}
\newcommand{\rme}{{\rm e}}

\begin{document}
\thispagestyle{empty}
\parskip=12pt
\raggedbottom

\def\mytoday#1{{ } \ifcase\month \or
 January\or February\or March\or April\or May\or June\or
 July\or August\or September\or October\or November\or December\fi
 \space \number\year}
\noindent
\hspace*{9cm} BUTP--95/15\\
\hspace*{9cm} COLO-HEP-362\\
\vspace*{0.5cm}
\begin{center}
{\LARGE Non--perturbative tests of the
           fixed point action for SU(3) gauge 
theory}\footnote{Work supported in part by Schweizerischer Nationalfonds,
NSF Grant PHY-9023257 
and U.~S. Department of Energy grant DE--FG02--92ER--40672}

\vspace{0.5cm}

Thomas DeGrand,
Anna Hasenfratz \\
Department of Physics \\
 University of Colorado,
Boulder CO 80309-390 

\vspace{.5cm}

Peter Hasenfratz, 
Ferenc Niedermayer\footnote{On leave from the Institute of Theoretical
Physics, E\"otv\"os University, Budapest} \\
Institute for Theoretical Physics \\
University of Bern \\
Sidlerstrasse 5, CH-3012 Bern, Switzerland

\vspace{0.5cm}

\mytoday \\ \vspace*{0.5cm}

\nopagebreak[4]

\begin{abstract}
In this paper (the second of a series) we extend our calculation of
a classical fixed point action for lattice $SU(3)$ pure gauge theory
to include gauge configurations with large fluctuations. The
action is parameterized in terms of closed loops of link variables.
We construct a few-parameter
approximation to the classical FP action which is valid for short 
correlation lengths.
We perform a scaling test of the action by computing the quantity
$G = L \sqrt{\sigma(L)}$ where the string tension $\sigma(L)$
is measured from
the torelon mass $\mu = L \sigma(L)$.  
We measure $G$ on lattices of fixed physical volume and varying
lattice spacing $a$ (which we define through the deconfinement temperature).
While the Wilson action shows scaling violations of about ten per cent,
the approximate fixed point action
scales within the statistical errors for $ 1/2 \ge aT_c \ge 1/6$. Similar
behaviour is found for the potential measured in a fixed physical volume.

\end{abstract}

\end{center}
\eject


\section{Introduction and summary}

This paper is the second in a series \cite{PAPER1} describing the construction
and testing of a fixed point (FP) action for lattice $SU(3)$ pure gauge
theory. In I we derived the general FP equations for the action and for
selected operators. We discussed their properties on the classical level
and in 1--loop perturbation theory. We solved the FP equation for
the action and the Polyakov loop on fields with small fluctuations. The
FP action is a perfect classical action on the lattice and, as we 
argued in I, it is perfect even in 1--loop
perturbation theory. It is the $\beta = \infty$ point of the renormalized
trajectory and, as such, will be the basic ingredient of future attempts
to construct perfect actions. This should be the ultimate goal in order
to reach the continuum limit without losing the battle 
against critical slowing
down, memory problems and other theoretical  or computational issues.

The aim of this paper is more modest. It is natural to expect, and this
expectation is supported by a numerical study of the non--linear 
$\sigma$--model \cite{HN}, that the FP action works much better than the
Wilson action even at small correlation lengths. In this paper we perform
a scaling test and make a comparison.

Scaling means that all physical dimensional
quantities show the same functional dependence on the gauge
coupling. It is different from asymptotic scaling where in
addition to scaling we require that this functional dependence be
described by the 2-loop $\Lambda$ parameter.
Asymptotic scaling properties
can be significantly improved by a suitable non-perturbative
redefinition of the bare coupling \cite{FirstDef,LM}. In this paper we are not
concerned with asymptotic scaling. Scaling properties can only  be
changed by modifying the lattice action.
A redefinition of the coupling constant used in the
lattice action during simulations
 will also shift  parameters in the lattice action
away from their naive free-field values.
Simulations done with these new actions may show better scaling properties.
An example of the use  of such actions
is the self-consistent tadpole-improvement
program for nonrelativistic QCD
which has been applied with great success to studies
of charm and bottom quark systems\cite{NRQCD}.

Scaling violations can be different for different quantities. The
ratio of certain observables might show scaling at smaller correlation
length than others. Scaling requires the universal behavior of {\it
all} observables. Of course, we cannot test all observables, but we can ask whether observables which show scaling violations when computed using
the Wilson action scale better when using a FP action.

The first step is to derive a parametrization for the FP action which
is sufficiently simple and approximates
well the FP action on rough configurations which typically enter
our simulations. This is a multistep procedure which we discuss in
some detail in Sections 2 and 3. At the end we obtained an 8 and a 
16 parameter approximation of the FP actions for RG transformation of
type I and type II, respectively. (These RG transformations were defined
and discussed in detail in I.)
 These parametrizations are constructed
in terms of two loops, the plaquette and the twisted perimeter-six loop.
Powers of the real and the square of the imaginary part of the loop--trace 
enter the 
action with coefficients  $c_1, c_2,...$ and $d_1, d_2,...$, respectively as
given in Tables \ref{tab:eightpar} and 
\ref{tab:sixteenpar}.

\begin{table*}[hbt]
\caption{Couplings of the few-parameter FP action for the RG
transformation of type I. The couplings d are zero.}
\label{tab:eightpar}
\begin{tabular*}{\textwidth}{@{}l@{\extracolsep{\fill}}lcccccc}
\hline
operator &  $c_1$ & $c_2$ & $c_3$ & $c_4$ \\
 \hline
$c_{plaq}$ &  .523  &  .0021      & .0053 & .0167   \\
$c_{6-link}$ &  .0597  &  .0054 &  .0051 & -.0006        \\
\hline
\end{tabular*}
\end{table*}

\begin{table*}[hbt]
\caption{Couplings of the few-parameter FP action for the RG
transformation of type II.}
\label{tab:sixteenpar}
\begin{tabular*}{\textwidth}{@{}l@{\extracolsep{\fill}}lccccccc}
\hline
operator &  $c_1$ & $c_2$ & $c_3$ & $c_4$ & $c_5$ \\
 \hline
$c_{plaq}$ &  .6092  &  .0478      & -.0470 & .0295 & -0.0038   \\
$c_{6-link}$ &  .0489  &  -.0146 &  .0393 & -.0216 & .0039       \\
\hline
operator &  $d_1$ & $d_2$ & $d_3$ & & \\
 \hline
$c_{plaq}$ &  .0921  &  -.1487      & .0397  & &  \\
$c_{6-link}$ &  .0042  &  -.0034 &  .0042    & &   \\
\hline
\end{tabular*}
\end{table*}

We pushed the scaling test to the extreme, going to configurations
with very large fluctuations (corresponding to Wilson $\beta = 5.1$).
This extreme situation might be interesting for finding the connection
with strong coupling expansions \cite{GA} but is, presumably, not
practical in standard applications. On very rough configurations it is
increasingly difficult to find a simple parametrization and, in
addition, the correlation length is so small that the signal disappears
rapidly.

In the scaling test which we performed with the 8 parameter action in
Table \ref{tab:eightpar}, the critical temperature $T_c$ was used 
to set the physical
scale. We determined the critical coupling constant $\beta_c(N_t)$ for
for $N_t = 2,3,4$ and $6$ (Table \ref{tab:betacrit}) and fixed 
the lattice spacing
at these coupling values as $a = 1/(T_c N_t)$. The corresponding
critical couplings for the Wilson action are known to  good 
precision \cite{FINGBERG}.

\begin{figure}[htb]
\begin{center}
\leavevmode
\epsfxsize=90mm
\epsfbox{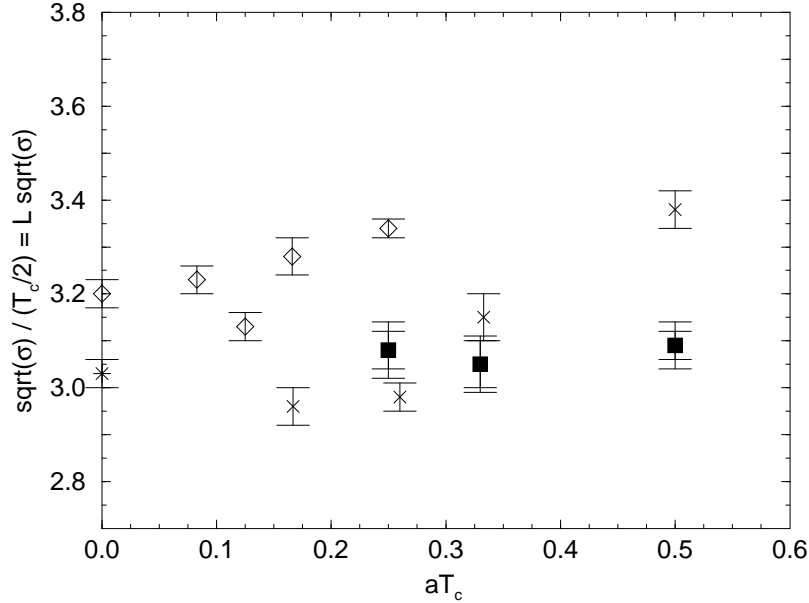}
\end{center}
\caption{Scaling test for the Wilson action (crosses) and FP action
 (squares); $T_c$ is defined in infinite
spatial volumes. 
The diamonds are  from zero-temperature simulations
by Ref. 8 and the diamond and the star at $a=0$ are extrapolations.
}
\label{fig:aspect_2a}
\end{figure}

Next we measured physical observables at couplings $\beta_c(N_t)$ in
fixed, finite physical volumes. This way we can avoid infinite volume
extrapolations.  For one scaling test we consider
the quantity
$G=L\sqrt{\sigma(L)}$ where $\sigma(L)$ is the string tension on
an $L^3$ volume computed from the exponential fall-off of the Polyakov
line correlator (torelon mass). Taking $L = r N_t$ where $r$ is some
conveniently chosen aspect ratio, the physical volume is kept fixed
at $V = (r/T_c)^3$. The ratio $G$ is independent of the bare coupling
(resolution) in the scaling (continuum) limit. Any variation of $G$
is due to lattice artifacts. Figure \ref{fig:aspect_2a}
shows $G(L)$ with $r = 2$
as a function of $aT_c$ for the Wilson and for the 8 parameter 
FP action.
The inner and outer error bars on the FP points show their
statistical uncertainty, and the combined uncertainty from statistics
and  in $\beta_c$.
The Wilson action shows a scaling violation of about $10\%$ between
$N_t = 2$ and $6$. No scaling violation above the statistical errors
is seen for the FP action. 

No simulations using the Wilson action at small lattice spacing precisely
match our scaling test using the torelon mass on fixed physical volumes.
The ones which come closest are the very recent ones of
Boyd. et.~al. \cite{BIELEFELD}. These authors present measurements
of the quantity $\sqrt{\sigma}/T_c$ at the critical couplings for
$N_t=4$, 6, 8, and 12, for which the string tension has been computed
via fits to  Wilson loops on large (zero temperature) lattices.
(They also present new measurements of critical couplings at $N_t=8$ and 12.)
The string tension from Wilson loops is an upper bound on $\sigma(L)$;
$\sigma(L)$ is reduced from $\sigma$ by the zero-point fluctuation term 
which for large $L$ in a string model is \cite{LSW}
\begin{equation}
\sigma(L) = \sigma - {\pi \over {3L^2}} + \dots .
\label{STRING}
\end{equation}

The diamonds in Fig. \ref{fig:aspect_2a} show $2 \sqrt{\sigma}/T_c$
from Ref. \cite{BIELEFELD}. The authors of that work present an extrapolation
to $a=0$, which we also display along with the $a=0$ limit of
$L\sqrt{\sigma(L)}$ at $r=2$ calculated from Eq. (\ref{STRING}).

Torelon masses at small lattice spacing have been presented in the
 literature, but they are not at couplings which are critical couplings
for deconfinement for integer $N_t$'s. Our attempts to interpolate them
into scaling tests which match our simulations generally resulted in
uselessly large uncertainties, and we do not show them here.

\begin{figure}[htb]
\begin{center}
\leavevmode
\epsfxsize=90mm
\epsfbox{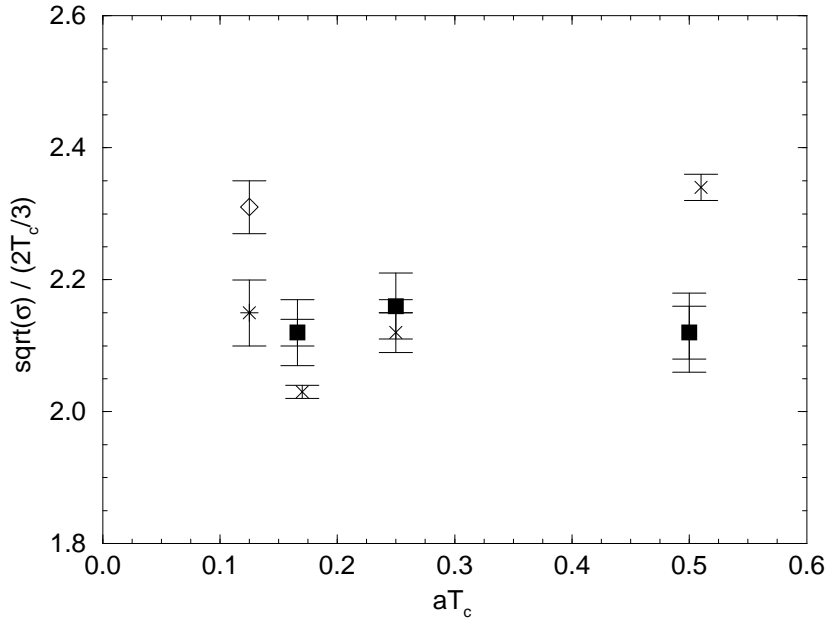}
\end{center}
\caption{Scaling test for the Wilson action (crosses) and FP action
 (squares) on aspect ratio 3/2 lattices. $T_c$ is defined in infinite
spatial volumes.
The diamond is a Wilson action simulation by us at $\beta=6.0$ and the star 
is an extrapolation using two-loop scaling to the correct (Ref. 8)
 $\beta_c=6.06$.
}
\label{fig:aspect_32a}
\end{figure}

Results are similar from  Fig. \ref{fig:aspect_32a} where the aspect ratio
$r = 3/2$ data are plotted. In  both cases the question remains 
unanswered, whether the constant observed with the FP action at very
low resolutions really agrees with the continuum value.


\begin{figure}[htb]
\begin{center}
\leavevmode
\epsfxsize=90mm
\epsfbox{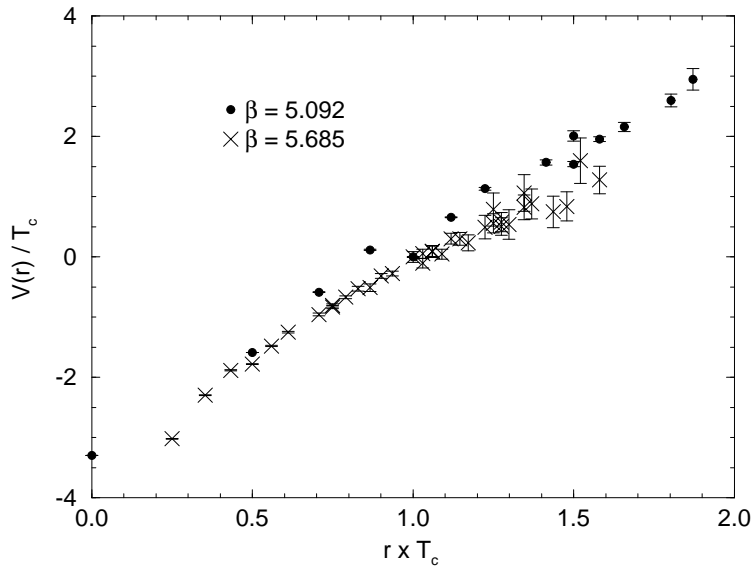}
\end{center}
\caption{  Potential $V(r)/T$ vs. $rT$ for the Wilson action at
$\beta_c(N_T=2)$ (filled circles) and $\beta_c(N_T=4)$ (crosses).  }
\label{fig:wilson24}
\end{figure}

\begin{figure}[htb]
\begin{center}
\leavevmode
\epsfxsize=90mm
\epsfbox{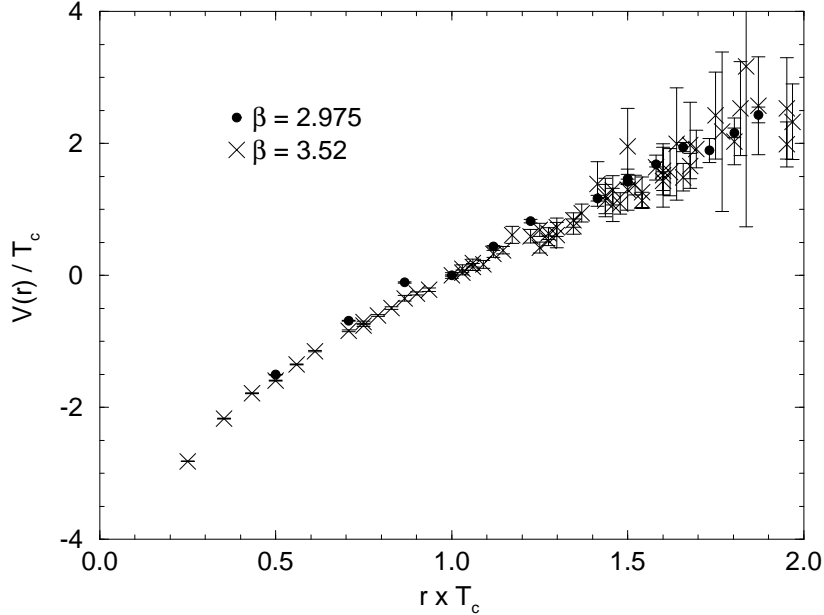}
\end{center}
\caption{ Potential $V(r)/T$ vs. $rT$ for the FP action at
$\beta_c(N_T=2)=2.975$ (filled circles) and $\beta_c(N_T=4)=3.52$
 (crosses).}
\label{fig:perfect24}
\end{figure}

Another scaling test is offered by the $q-\bar q$ potential $V(r;L)$ in
a volume $L^3$ where $L$ is chosen here to be $L=2N_t$. Unfortunately,
this quantity is contaminated by cut--off effects for $r \sim a$ 
coming from non--perfect ${\it sources}$. The FP Polyakov loop has not
been constructed yet beyond the quadratic approximation (paper I),
so we measured simple Polyakov loop correlations. The different
behaviour of the potentials obtained with the Wilson and FP actions
is well demonstrated by Figs. \ref{fig:wilson24} and 
\ref{fig:perfect24}, 
nevertheless.

In the potential of the FP action the rotational symmetry violation at
large $r$ drops below the statistical errors even at $\beta_c(N_t)=2$
but remains clearly present for the Wilson action. (Compare, for example,
the predictions at $r=1.5/T_c$ coming from $\vec{r}=(0,0,3)$ and 
$\vec{r}=(1,2,2)$ at  $\beta_c(N_t)=2$.) The non--physical additive constant
in the potential was fixed by the convention $V(r=1/T_c) = 0$. The lack
of scaling for the Wilson action is evident, while the long distance 
behaviour of the potential is consistent with scaling for the 
8 parameter FP action.

The outline of the paper is as follows: In Section 2 we describe the
construction of a classical FP action which is appropriate for gauge
configurations whose fluctuations are not small. In Section 3
we give an approximate simple parametrization of the FP action which
represents the FP on gauge field configurations which are typical
at the correlation lengths of our simulation. Finally, in Section 4
we give some details of the scaling tests using this action.

\section{Numerical analysis of the FP equation and the problem
of parametrization}

We consider an SU(N) pure gauge theory\footnote{The actual numerical
analysis and simulations in this paper were done for SU(3). The
equations are written for general N if not indicated otherwise.} on
the lattice and consider the RG transformation
\bee
\rme^{-\beta' S'(V)} = \int DU
\exp\left\{ -\beta \left( S(U)+T(U,V)\right) \right\}  ,        \label{1}
\ee
where $U$ is the original link variable, $V$ is the blocked link
variable and $T(U,V)$ is the blocking kernel that defines the 
transformation. At $\beta = \infty$ the transformation becomes
a steepest decent relation
\bee
S^{FP}(V)=\min_{ \{U\} } \left( S^{FP}(U) +T(U,V)\right),  \label{2}
\ee
where
\bee
T(U,V)=-{\kappa\over N}
\sum_{n_B,\mu} \left[ {\rm Re Tr} \left( V_{\mu}(n_B)
Q_{\mu}^{\dagger}(n_B)\right) -
\max_{W} \left\{ {\rm Re Tr} (W Q^{\dagger}) \right\} \right].
                                              \label{3}
\ee
In Eq.~(\ref{3}) $W \in {\rm SU}(N)$ and the $N \times N$ complex 
matrix $Q_{\mu}(n_B)$ is the block average.
Its form depends on the block transformation. For the two different
block transformations (type I and type II) which are considered here
$Q$ is defined in Figs.~5 and 6 in I. For further details we refer
also to I.

In I we solved Eq.~(\ref{2}) up to cubic (quadratic) order in the
vector potentials for type I (type II) block transformation. If the
configuration ${V}$ has large fluctuations one has to use numerical
methods to solve the FP equation.

For numerical simulations we need a parametrization for the FP action
which is sufficiently simple to make the computational overhead
acceptable. It turns out, however, that the process of solving
Eq.~(\ref{2}) requires, as an intermediate step, a more general
parametrization. This is the problem we discuss first.

\subsection{Parametrization of actions}

Our parametrization is based on powers of the traces of the 
loop products $U_{\it C} =  \Pi_{\it C} U_\mu(n)$
where $\it C$ is an arbitrary closed path
\begin{eqnarray}
S(U) = {1 \over N} \sum_{\it C} ( c_1({\it C})(N-
{\rm ReTr}(U_{\it C})) & +&
                       c_2({\it C})(N-{\rm ReTr}(U_{\it C}))^2+ ...
\nonumber  \\
                    +  d_2({\it C}) ({\rm ImTr}(U_{\it C}))^2 & + &
               d_4({\it C}) ({\rm ImTr}(U_{\it C}))^4 \; \; ...).
\label{4}
\end{eqnarray}

We  also checked other possibilities. The success of a parametrization
of the $\sigma$-model and the form of exact classical solutions in
both models suggest  building gauge invariant quantities from
$U_{\it C}$ by using the angles after diagonalization,
$U_{\it C}^{diag} = (e^{i\theta_1}, e^{i\theta_2}, e^{i\theta_3})$
for $N = 3$ \cite{Manton}. For SU(3) gauge theory this type of parametrization
did not work better than the one in  Eq.~(\ref{4}) and was, therefore,
abandoned. We also tried to include the product of traces of different
loops. It is quite difficult to keep track of the large number of
topological possibilities and, in those cases we checked, the 
parametrization did not become better. The following results are
based on the parametrization in  Eq.~(\ref{4}).

Using optimized parameters in the block transformation the quadratic
and cubic part of the FP action is short ranged (paper I). So,
we decided to restrict the set of loops ${\it C}$ to the 28 topologically
different loops that are length 8 or less
and fit into a $2^4$ hypercube. They are tabulated in Table \ref{tab:loops}.
Only 7 (15) of the 28 are independent on the quadratic (cubic) level.
For the RG transformation of type I the FP action is known up to cubic
order analytically which can be used to fix 15 of the $c_1({\it C})$
coefficients by a $\chi^2$ fit. We decided to keep only 12 operators  since
the fit did not improve further. For the RG transformation of type II
7  $c_1({\it C})$ couplings were fixed this way since the analytic cubic
result is not available.  Table \ref{tab:couplingsI} and
Table \ref{tab:couplingsII} contain the
$c_1({\it C})$ coefficients obtained. We shall denote the action where
these couplings are kept only by $S_q^{FP}$.

\begin{table*}[hbt]
\newlength{\digitwidth} \settowidth{\digitwidth}{\rm 0}
\catcode`?=\active \def?{\kern\digitwidth}
\setlength{\tabcolsep}{1.5pc}
\caption{Loops used in the construction of the FP action.
One particular orientation is shown; all possible orientations
are included in the actual operators.}
\label{tab:loops}
\begin{tabular*}{\textwidth}{@{}l@{\extracolsep{\fill}}lcccccc}
\hline
$label$ &  path \\
 \hline
1  &x,y,-x,-y \\
2  &x,y,y,-x,-y,-y \\
3  &x,y,z,-y,-x,-z \\
4  &x,y,z,-x,-y,-z \\
5  &  x,  y,  x,  y,  -x,  -y,  -x,  -y \\
6  &  x,  y,  x,  y,  -x,  -x,  -y,  -y \\
7  &  x,  y,  x,  z,  -y,  -x,  -x,  -z \\
8  &  x,  y,  x,  z,  -x,  -z,  -x,  -y \\
9  &  x,  y,  x,  z,  -x,  -y,  -x,  -z \\
10  &  x,  y,  x,  z,  -x,  -x,  -z,  -y \\
11  &  x,  y,  x,  -y,  -x,  y,  -x,  -y \\
12  &  x,  y,  y,  x,  -y,  -x,  -x,  -y \\
13  &  x,  y,  y,  z,  -y,  -z,  -x,  -y \\
14  &  x,  y,  y,  z,  -y,  -y,  -x,  -z \\
15  &  x,  y,  y,  z,  -y,  -x,  -z,  -y \\
16  &  x,  y,  y,  z,  -x,  -y,  -y,  -z \\
17  &  x,  y,  y,  -x,  z,  -y,  -y,  -z \\
18  &  x,  y,  y,  -x,  -x,  -y,  -y,  x \\
19  &  x,  y,  z,  t,  -z,  -t,  -x,  -y \\
20  &  x,  y,  z,  t,  -z,  -y,  -x,  -t \\
21  &  x,  y,  z,  t,  -z,  -x,  -t,  -y \\
22  &  x,  y,  z,  t,  -z,  -x,  -y,  -t \\
23  &  x,  y,  z,  t,  -y,  -x,  -t,  -z \\
24  &  x,  y,  z,  t,  -y,  -x,  -z,  -t \\
25  &  x,  y,  z,  t,  -x,  -y,  -z,  -t \\
26  &  x,  y,  z,  -y,  -z,  y,  -x,  -y \\
27  &  x,  y,  z,  -y,  -x,  y,  -z,  -y \\
28  &  x,  y,  -x,  -y,  x,  y,  -x,  -y \\
\hline
\end{tabular*}
\end{table*}

\begin{table*}[hbt]
\setlength{\tabcolsep}{1.5pc}
\caption{Couplings of the  fixed point action of Type I
given in terms of
loops defined in Table 3. The quadratic couplings are labelled by $c_1$,
and the coefficients of higher powers of $(3- {\rm Tr}U(C)$
from the fit described in Sec. 2.2 are also shown.}
\label{tab:couplingsI}
\begin{tabular*}{\textwidth}{@{}l@{\extracolsep{\fill}}lcccccc}
\hline
loop & $c_1$ & $c_2$ & $c_3$  &$c_4$ \\
   1 &   .6744  & .1067  &   -.1072 & .0186 \\
   2 &  -.02 & .0424 & -.0304 & .0055 \\
   3 &  .012  & .0361  & -.0228 & .00436 \\
   4 &  .005 & .1132  &  -.0632 & .00937 \\
   5 &  -.0031 & .0765 & -.0562 & .0109 \\
   8 &  -.0035 & .00525 & -.00418 & .00076 \\
   9 &  .0027 & .00402 & -.00642 & .00130 \\
   15 &  -.0024 & -.0356 & .0187 & -.00268 \\
   16 &  .0013 & -.0023  &-.00373 & .00132 \\
   20 &  .003 & .0193 & -.0119 & .00211 \\
   23 &  .0032 & .0187 & -.0128 & .0024 \\
   25  & .0024 & .0126 & -.0115 & .0021 \\
\hline
\end{tabular*}
\end{table*}

\begin{table*}[hbt]
\setlength{\tabcolsep}{1.5pc}
\caption{Couplings of the  fixed point action of Type II,
labeled as in Table 4 .}
\label{tab:couplingsII}
\begin{tabular*}{\textwidth}{@{}l@{\extracolsep{\fill}}lcccccc}
\hline
loop & $c_1$ & $c_2$ & $c_3$  &$c_4$ \\
   1 &   .5454  & -.0837  &  0.0459 & -.0076 \\
   2 &   .0094 & -.0335 & .0161 & -.0026 \\
   3 &   .0  & .0639  & -.0403 & .0076 \\
   4 &   .0419 & .0594  &  -.0485 & .0091 \\
   5 &   .0 & .0088 & -.0060 &  .0010 \\
   8 &   .0 & -.0022 &  .0008 & -.0001 \\
   9 &   .0019 & .0033 & -.0019 & .0003 \\
   17 & -.0011 & .0035 & -.0015 & .0002 \\
   18 & -.0041 & -.0081  &.0040 & -.0006 \\
   20 &  .0 & -.0109 &  .0074 & -.0011 \\
   23 &  .0 & -.0213 &  .0111 & -.0014 \\
   25  & .0043 & .0054 & -.0039 & .0007\\
\hline
\end{tabular*}
\end{table*}

\subsection{Numerical minimization}

To go beyond leading order we solved the fixed
point equation Eq.~(\ref{2}) numerically.
The procedure is the following:
First we create  an arbitrary SU(3) configuration  \{V\}.
That serves as a coarse
configuration. Our goal is to find a fine configuration \{U\} according
to Eq.~(\ref{2}). Since we do not know $S^{FP}$ (and even if we knew it,
it presumably would be too complicated to use in a numerical minimization),
we approximate $S^{FP}$ with a simpler action $S_0$. We choose
$S_0$ such that it is sufficiently
simple for minimization but close to the fixed point action.
The minimization of  Eq.~(\ref{2})
gives the fine configuration $\{U_0\}$ as a minimum
\bee
S^\prime(V)=\min_{U}(S_0(U)+T(U,V))= S_0(U_0)+T(U_0,V).
\label{5}
\ee


If $S_0$ is close to the fixed point action $S^{FP}$, than $U_0$ is close
to $\{U^{FP}\}$ and up to quadratic corrections
\begin{eqnarray}
 S^{FP}(V)= & S^{FP}(U_0)+T(U_0,V)  \nonumber \\
       =& S^{FP}(U_0) + (S^\prime(V)- S_0(U_0)).
\label{6}
\end{eqnarray}
This equation can be used to calculate the value of the fixed point
action on the course configuration \{V\} if $S^{FP}$ is known on the
fine configuration $\{U_0\}$.

One might think we got nowhere this way but this is not so. One 
observes that the minimizing configuration $U^{FP}(V)$ (which is close
to  $U_0(V)$) has much smaller fluctuations than the configuration ${V}$.
As  Eq.~(\ref{3}) shows, $T(U^{FP}(V),V)$ is non-negative and so
 Eq.~(\ref{2}) implies that $S^{FP}(V) \ge S^{FP}(U^{FP}(V))$. From this 
inequality it follows that the action density on the fine lattice is
at least ${1 / 2^4}$ times smaller than on the coarse one.
Actually, the observed ratio is even smaller, it lies between
 ${1} / {30}$ and  ${1} / {50}$ on the configurations we
considered\footnote{In order to appreciate how small is this ratio
let us remark that the Wilson action density 
$(3 - {\rm ReTr}U_p)$ 
is reduced by a factor of $\sim 2$ as $\beta$ is changed
from 5.0 to 10.0.}. On the configurations $U_0(V)$ which have very
small fluctuations the FP action can be replaced by  $S_q^{FP}$
defined at the end of Section 2.1  \footnote {One might and we
did include further terms in order to cope with the small
corrections which arise when $V$ is very coarse and so the
fluctuations of $U_0(V)$ are not so small anymore. In order
to keep the discussion straight we do not follow this point
further.}. Eq.~(\ref{6}) gives then the value of the FP action
on the configurations $V$.

\begin{figure}[htb]
\begin{center}
\leavevmode
\epsfxsize=90mm
\epsfbox{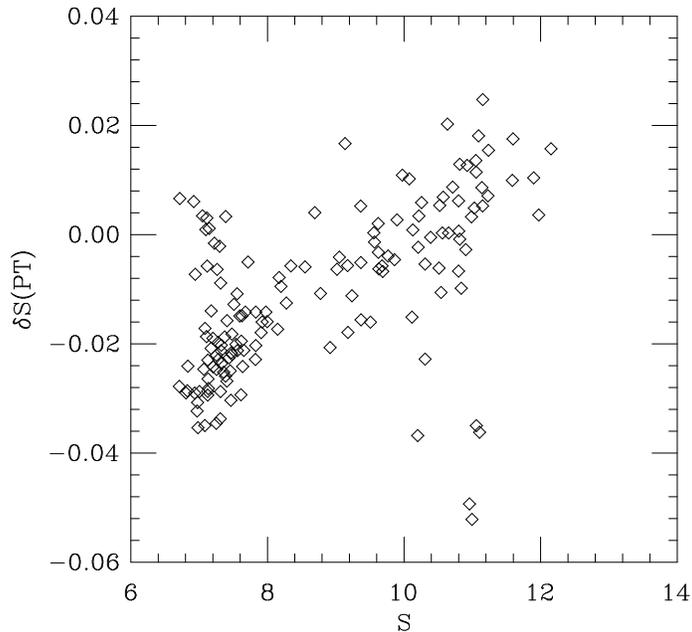}
\end{center}
\caption{ The perturbative corrections
$\delta S_{PT}=S^{FP}(U_0)-S_0(U_0)$ for a set of configurations
using the type I RGT.
}
\label{fig:pert}
\end{figure}

The procedure described above could fail if $S_0$ in the original
minimization was not close to the FP action. We found that by 
carefully adjusting the couplings of $S_0$ one can make the 
difference $S^{FP}(U_0)-S_0(U_0)$, that measures this correction,
small. After a certain amount of trial and error we were able
to reduce these ``perturbative'' corrections to less than 0.4\%
and 0.3\% of the action for the type I and type II RG transformations,
respectively.
In fig.~\ref{fig:pert} we illustrate this fact for the type I RGT.

In the case of the RG transformation of type I we generated 
about 400 configurations on  $2^4$ and $4^4$ coarse lattices using
the Wilson action, with
coupling $\beta$ ranging from 5.0 to 7.0. For type II we created
more than 700 configurations in the coupling constant range
5.0 to 50.0. Following the steps of the discussion above
we associated to each configuration a real number $S^{FP}(V)$,
i.e. the value of the FP action on the configuration.

\begin{figure}[htb]
\begin{center}
\leavevmode
\epsfxsize=90mm
\epsfbox{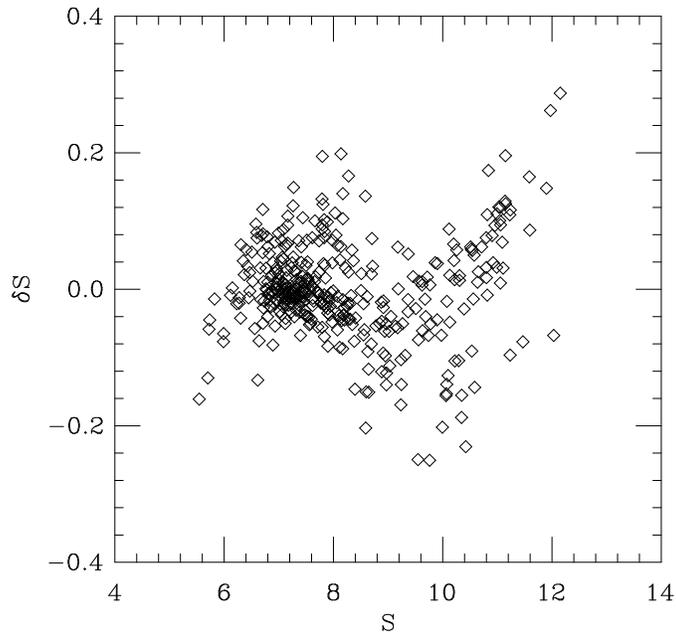}
\end{center}
\caption[]{  $\delta S$ (the difference between the two
sides of Eq.~(\ref{2})) vs. $S_{min}$ for  the many parameter
fit of Table 4 for the type I RGT.  }
\label{fig:multipar}
\end{figure}

Our next task is to find a parametrization for  $S^{FP}$ which
reproduces the action values to an acceptable accuracy and, at
the same time, is sufficiently simple to be used in numerical
simulations. This problem will be discussed in Section 3. Beyond
that we would like to make the information stored in the 
generated configurations and action values available to those
who might want to create their own simple parametrization for
simulations. For this purpose we parametrized $S^{FP}$ as in
 Eq.~(\ref{4}) keeping a large number of loops from the set in 
Table \ref{tab:loops}. The couplings are presented in 
Table \ref{tab:couplingsI} and Table \ref{tab:couplingsII}
for the RG transformations of type I and type II, respectively.
Of course, these complicated actions are not practical to
be used in simulations. They are used only to encode 
information which would be difficult to present otherwise.
The actions in Table \ref{tab:couplingsI} and 
Table \ref{tab:couplingsII} satisfy the FP equation 
Eq.~(\ref{2}) on our configurations to a reasonable precision.
The difference between the left and right hand sides of
Eq.~(\ref{2}) is no more than 2\% of the action, and
in most of the cases the difference is much less.
The quality of the fit is illustrated in fig.~\ref{fig:multipar}.

\subsection{Technical remarks on numerical minimization}

In order to avoid programming errors two codes were written
independently (this applies for all the essential parts of
the numerical analysis presented in this paper). In both cases
the lattice was swept through repeatedly, making a random local
change and accepting the change if it lowered the action. The local
change consisted a random rotation on a fine link, or on two
neighbouring fine links at the same time. The rotations were
parametrized and the action was interpolated quadratically in the
parameters to find a set of values which minimized the action
in this local few-parameter space. The configuration was overwritten
 only after checking that the action was really lowered.

In order to speed up the minimization process it was useful to
begin with a fine configuration close to the final answer. One
strategy for constructing
 such a configuration was to drop the
${\rm Re Tr} WQ^\dagger$ normalizing term from $T(U,V)$ during the
minimization 
and to relax for 20 to 50 sweeps. This typically
brought the action to within ten per cent of its minimum value. Then
20 to 50 sweeps of the full minimization were required to achieve
convergence at the fourth digit of accuracy.
Another method was to build a fine starting configuration using
the connecting matrix $Z$ (Eq.(25) in I) valid on configurations
with small fluctuations. This method, which required temporary gauge
fixing, worked suprisingly well even on rough configurations.

\section{The few-parameter FP action}

For numerical simulation we need a parametrization which is
simple and approximates the FP action sufficiently well on those
configurations which are typical in the actual calculation. In order
to find such a parametrization we used coarse configurations
generated by the Wilson action between $\beta = 5.2$ and 6.0 and
fitted $S^{FP}(V)$ with  powers of the traces of only 2 loops.

The two loops we used in our few-parameter fit were the plaquette
and loop \#4 in Table \ref{tab:loops}, the twisted 6 link operator
(x,y,z,-x,-y,-z). For RG transformation of type I we kept $c_i,
i=1,...,4$ and put $d_i = 0$. The fit for the 8 parameters are listed
in Table \ref{tab:eightpar}.
For RG transformation of type II we parametrized with $c_i, i=1,...,5$
and $d_i, i = 1,...,3$. The corresponding 16 fitted parameters are
given in Table \ref{tab:sixteenpar}.

\begin{figure}[htb]
\begin{center}
\vskip 10mm
\leavevmode
\epsfxsize=90mm
\epsfbox{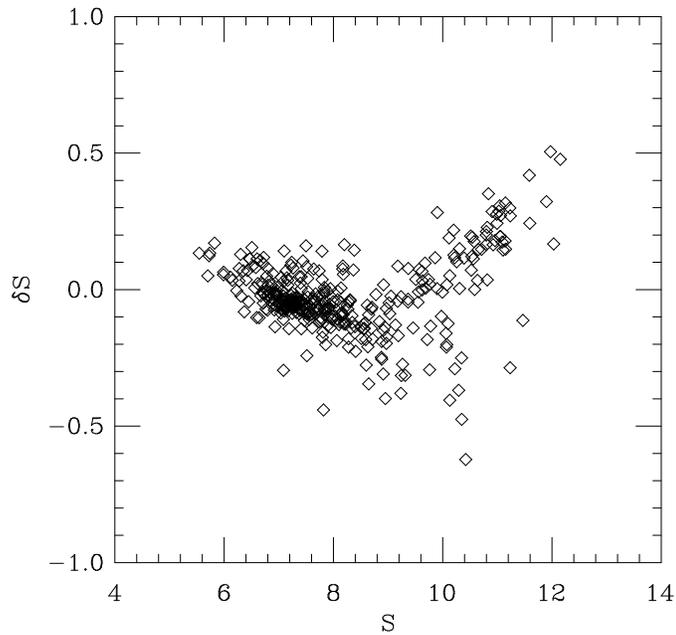}
\vskip 10mm
\end{center}
\caption{$\delta S= S^{fit}(V)-S^{FP}(V)$ vs. $S^{FP}$ for  
the 8 parameter fit.  }
\label{fig:fewpar}
\end{figure}

The accuracy of the fit  is illustrated for the 8 parameter action
in Fig. \ref{fig:fewpar} where $\delta S= S^{fit}(V)-S^{FP}(V)$
is plotted as a function of
$S^{FP}$. The quality of the fit deteriorates for smaller correlation
lengths (larger $S^{FP}$) but remains within the 5\% range.

\begin{figure}[htb]
\begin{center}
\vskip 10mm
\leavevmode
\epsfxsize=90mm
\epsfbox{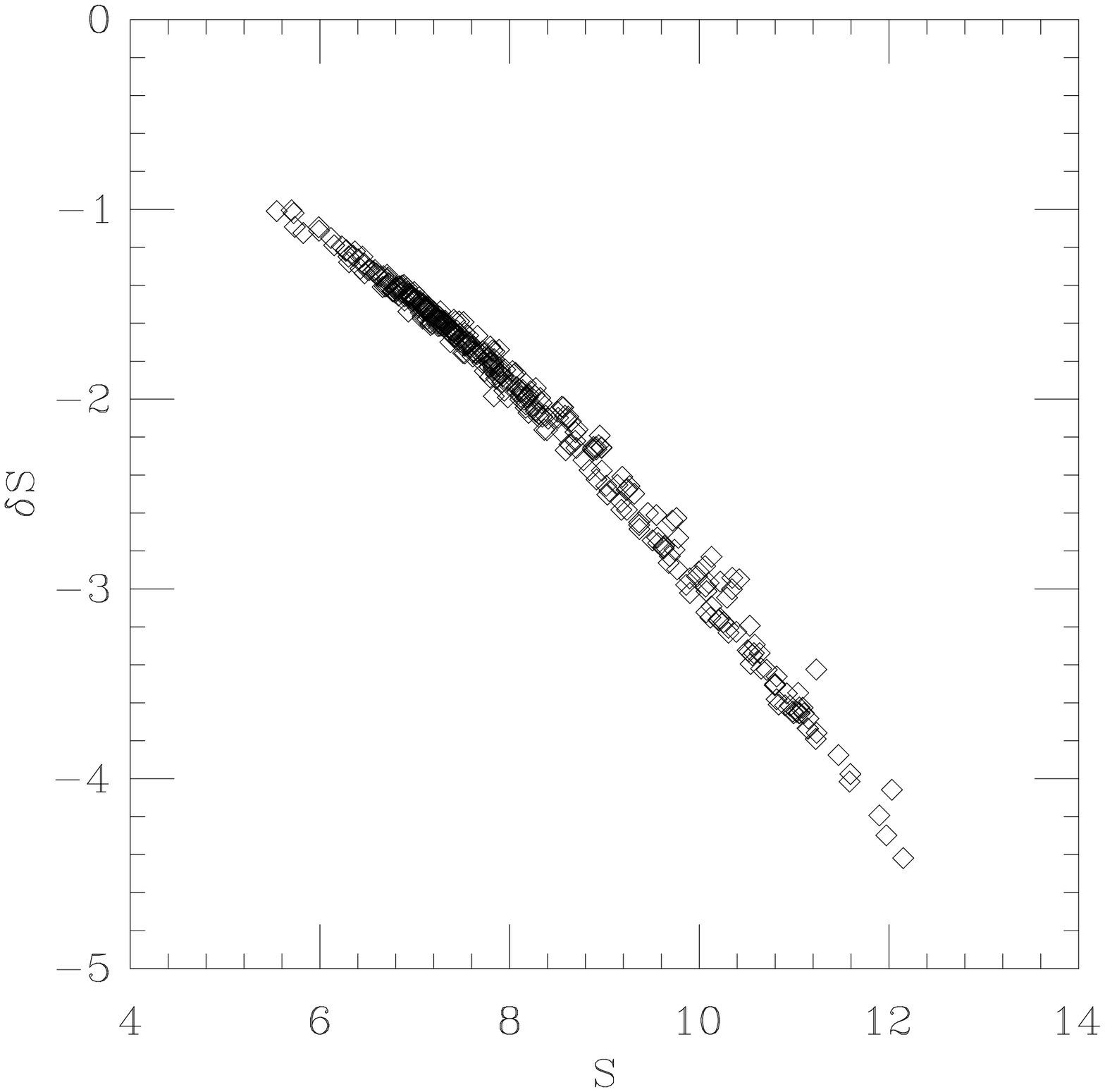}
\vskip 10mm
\end{center}
\caption{  $\delta S$ vs. $S_{min}$ using {\it only} the quadratic
coefficients of the action.
}
\label{fig:quad_only}
\end{figure}

As is evident from Tables \ref{tab:eightpar} and \ref{tab:sixteenpar},
the FP action is dominated by the plaquette term with a small
 contribution of the 6 link operator in the lowest power.
 Can we neglect higher powers of loops
 altogether? To illustrate the importance of these
small coefficients, in
Fig  \ref{fig:quad_only} we plot $\delta S= S^{fit}(V)-S^{FP}(V)$ 
calculated with {\it only} the
quadratic $c_1$ coefficients. It is obvious that  the seemingly small
$c_i, i = 2,3,4$ terms play a very important role in fitting the FP
action.

The parameters in Tables \ref{tab:eightpar} and \ref{tab:sixteenpar}
were chosen to approximate the FP action on configurations typical
of a correlation length range occurring in $\beta_{Wilson} \sim (5.2,6.0)$.
If we wanted for some reason to
perform simulations at $10^{-3}$ fm, we would use a parameterization
of the FP action which was valid at the small lattice spacing.

\section{Simulations with approximate FP actions}

The FP action is the classically perfect lattice action. As we argued
in I, it is perfect even in 1-loop perturbation theory. These features
make it plausible to expect that the FP action will have improved
properties even at small correlation lengths. We present a detailed
scaling test for the FP action of the RG transformation of type I
using the approximate parametrization in Table \ref{tab:eightpar}.

\subsection{Simulation algorithms}

We have adapted two algorithms for simulation. First, we have written a mixed Metropolis/overrelaxation algorithm which
acts on SU(2) subgroups, in complete analogy with a standard Wilson
action code. No special optimization was done. The program runs with
8 parameters about a factor of 5 slower than with 1 parameter 
corresponding to the Wilson action. This latter is a factor of 1.4 slower
than a highly optimized Wilson code. There are three reasons
for the loss of speed. The first
 is the extra loops needed by the new action.
The second is that because the action is not linear in a link variable,
one cannot sum the links in the staples ahead of time, but must do so for
each new update. Finally, again because the action is not linear in the link
variables,
we have not been able to invent a microcanonical overrelaxation step,
and after each overrelaxation step
we must recompute the action and perform a Metropolis accept/reject
step. The recomputation is unnecessary for the Wilson action. We have not
precomputed any parts of the action for greater speed; the program
is designed for arbitrary actions.

All our running on serial machines has used this algorithm. We typically
do a mix of four overrelaxation updates on each of three $SU(2)$ subgroups
on each link, followed by a one-hit Metropolis
update (this is a ``sweep'' as used below), mimicking a conventional updating pattern
for the Wilson action. 
We have not done long enough runs to have measured
reliable simulation
autocorrelation times. They appear to be about five sweeps
for the plaquette, measured near the $N_t=4$ critical coupling.
The autocorrelation time was negligible for the correlators of Polyakov
loops. We typically measure all quantities once per sweep.
Of course, the whole point of the perfect action program is to be able to run
on small lattices with small correlation lengths where critical slowing
down is absent.

The overrelaxed/Metropolis algorithm is rather inconvenient for
parallel machines. The action does not parallelize
on two checkerboards due to the presence of the six-link term, and
 because the action includes more than
 loops in the fundamental representation,
much temporary storage is needed. Neither problem is insurmountable, but we
avoided them by
using the Hybrid Monte Carlo algorithm \cite{HMC}. The HMC code
we have written for the eight-parameter FP action is about a factor of seven
slower than a highly optimized Wilson action HMC code \cite{MILC}.
The difference in speed is easily understood as coming entirely from the
many extra loop terms in the action.

\subsection{The method of testing the FP action}

When scaling holds the ratio of two physical masses is
independent of the lattice resolution, which can be changed by tuning
the coupling $\beta$. One of the physical masses we took is the critical
temperature $T_c$ which is used to set the physical scale. The lattice
spacing $a$
and $T_c$ are related in the usual way, by finding the critical coupling
$\beta_c(N_t)$ at which the confinement-deconfinement transition occurs
on a lattice of size $N_s^3 \times N_t$ (with $N_s \gg N_t)$. All our
additional simulations were done at the discrete $\beta$ values 
$\beta_c(N_t)$, where the lattice spacing is already fixed and given by 
$a = 1/(T_c N_t)$.

For the other physical mass we measured the torelon mass. A somewhat less
quantitative test is offered by the potential which we also measured.

In the first case the scaling test is the quantity 
 $G=L\sqrt{\sigma(L)}$ where $\sigma(L)$ is the string tension on
an $L^3$ volume measured through the torelon mass. We performed the
numerical experiment at two different aspect ratios $L/N_t = 2$ and 3/2.

In the second case we studied the potential $V(r;L)$ in a volume $L^3$,
where $L$ was chosen to be $L = 2N_t$. This test is less quantitative,
since, rather than using FP Polyakov loops which we have not yet
constructed beyond the quadratic approximation (paper I), the 
potential was extracted from simple Polyakov loop correlations.
This is correct only for $r>>a$ where it makes a negligible error
that the Polyakov loops create sources whose distance is not exactly $r$.

\subsection{Critical couplings for deconfinement}

 We found that the Columbia group's definition \cite{COL,DOUG}
of $\beta_c$ is the easiest to use
on small volumes.  Specifically, we make simulations of up to several
tens of thousands of
sweeps, recording the value of the real and imaginary parts of the Polyakov
loop averaged over the lattice. We form the angle $\theta$ from
$\tan \theta ={\rm Im} P/ {\rm Re} P$ and measure the fraction of the
 time $f_{20}$ that $\theta$
lies within $\pm 20$ degrees of one of the ordered orientations
expected for the Polyakov loop in the high temperature phase. We
define $\beta_c$ as the place where the deconfinement
fraction 
\begin{equation}
f_D = {3 \over 2} f_{20} - {1 \over 2}
\end{equation}
equals fifty per cent.
Most of our simulations are done on lattices of spatial size $L$ between
2 or 3 times $N_t$. On these sizes close to $\beta_c$ the lifetimes of
the various $Z_3$ vacua average a few hundred sweeps. As we will 
see, it is necessary to measure $\beta_c$ to within $\pm 0.005$ to make
meaningful scaling tests. We found alternate diagnostics such as histograms
of the average Polyakov loop modulus just less precise than $f_D$ measurements.

Infinite volume $\beta_c$ values are found by extrapolating
finite volume $\beta_c$ values linearly in 1/volume and also by looking
at crossings in a plot of Binder cumulants, which for us is
$B = 1 - \langle L^4 \rangle /(3 \langle L^2 \rangle^2) $,
where $L^2$ is the absolute square of the volume-averaged
Polyakov loop on a configuration, $L^4= (L^2)^2$, and $\langle \rangle$
is an average over a simulation run.

\begin{figure}[htb]
\begin{center}
\vskip 10mm
\leavevmode
\epsfxsize=90mm
\epsfbox{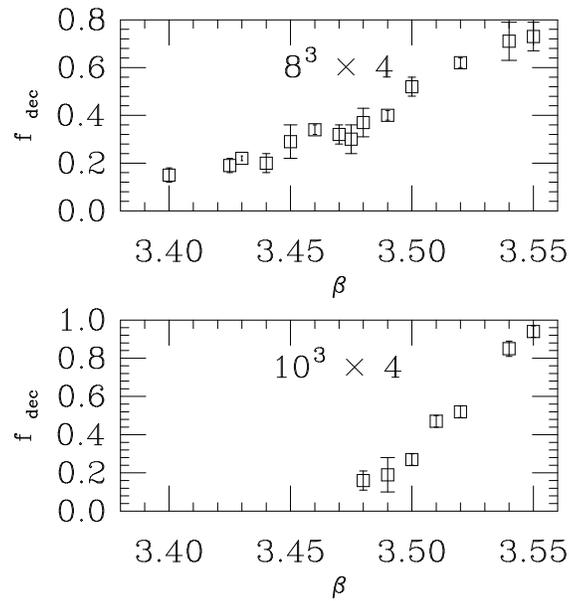}
\vskip 10mm
\end{center}
\caption{ Deconfinement fractions for the eight-parameter FP action
for $N_t=4$ lattices.}
\label{fig:examples}
\end{figure}

\begin{figure}[htb]
\begin{center}
\vskip 10mm
\leavevmode
\epsfxsize=90mm
\epsfbox{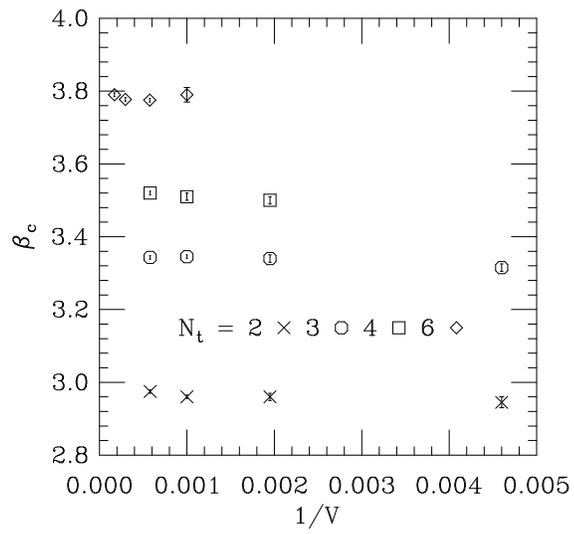}
\vskip 10mm
\end{center}
\caption{Volume dependence of critical couplings for 
the eight parameter FP
action, and their extrapolations to infinite volume.}
\label{fig:beta_c_vs_nt}
\end{figure}

\begin{figure}[htb]
\begin{center}
\vskip 10mm
\leavevmode
\epsfxsize=90mm
\epsfbox{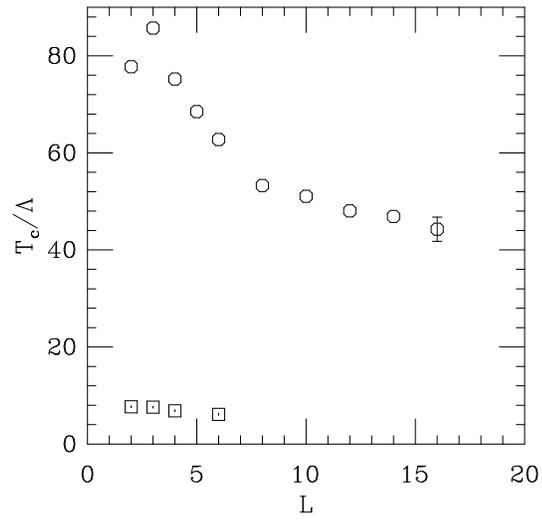}
\vskip 10mm
\end{center}
\caption{$T_c/\Lambda$ for the Wilson (octagons) and  FP (squares)
actions.
}
\label{fig:tc}
\end{figure}

We have measured critical couplings on lattices with $N_t=2$, 3, 4, and 6.
The spatial volumes and critical couplings
are shown in Table \ref{tab:betacrit}. 
Examples of plots of deconfinement fractions are shown 
in Fig. \ref{fig:examples}.
A plot of $\beta_c$ vs. 1/volume is shown in Fig. \ref{fig:beta_c_vs_nt}.
Although the FP action is not designed to improve asymptotic scaling,
just scaling, we can still see if asymptotic scaling obtains.
Using the original coupling in the action, we compute $T_c/\Lambda$
from the two loop formula from our data, and record the results in Table
\ref{tab:betacrit} and Fig. \ref{fig:tc}. The eight parameter 
FP action  not
only shows asymptotic scaling within twenty per cent
from $N_T=2$ to 6 but in addition the value of its
$\Lambda$ parameter is about a factor of ten larger than the Wilson
one and therefore much closer in value to the continuum $\Lambda$
parameters.

\begin{table*}[hbt]
\setlength{\tabcolsep}{1.5pc}
\caption{Critical couplings at finite volume and extrapolated to
infinite volume for the FP action with parameters in Table~1.}
\label{tab:betacrit}
\begin{tabular*}{\textwidth}{@{}l@{\extracolsep{\fill}}lcccccc}
\hline
volume   & $N_t=2$   & $N_t=3$ & $N_t=4$ & $N_t=6$ \\
$4^3$    & 2.89(1)     &    &    &    \\
$6^3$    &  2.945(15)  & 3.320(15)   &    &    \\
$8^3$    &  2.96(1)    &  3.34(1)  &  3.50(1)  &    \\
$10^3$   & 2.960(5)    &  3.34(1)  &  3.51(1)  &  3.78(2)  \\
$12^3$   &2.975(5)     &  3.343(5) &  3.520(5)  & 3.775(5)    \\
$15^3$   &             &           &            & 3.777(5)   \\
$18^3$   &             &           &            & 3.790(5)   \\
infinite &  2.976(5)   &  3.345(5) &  3.520(5)  &  3.790(6)  \\
\hline
$T_c/\Lambda$   & 7.68   & 7.61   & 6.87   &  6.13  \\
\hline
\end{tabular*}
\end{table*}

\subsection{Torelon masses}

The string tension $\sigma$ is measured through the correlator of pairs of
Polyakov loops (or more complicated objects; see below): on a lattice
of transverse size $L$ the correlator of two Polyakov loops averaged 
over transverse separations and separated
a longitudinal distance $z$ is
\begin{equation}
C(z) = \sum_{r_\perp} {\rm Re} \langle P_j(r_\perp,z)P_j^\dagger(0,0) \rangle
 \simeq \exp(-\mu z),
\end{equation}
where $\mu$ is the so-called torelon mass. On an infinite lattice
$\mu = \sigma L$ and we will make the same definition on a finite lattice.

A problem with Polyakov loop correlators is that the signal to noise
ratio rises exponentially with the string tension, $L$, and $z$:
a simple calculation shows that with $N$ measurements
\begin{equation}
{\rm Signal/Noise} \simeq \sqrt{N}\exp(-\sigma Lz).
\label{SNratio}
\end{equation}
Since as $\beta$ falls, $\sigma$ rises, this means that at small $\beta$
it is difficult to observe a signal unless $z$ is very small. However, at
small $z$ the signal may be contaminated by excited states, and it may
be impossible to go to large enough $z$ to be certain that one is seeing
the asymptotic behavior of the signal before it disappears in the noise.

We are primarily interested in lattice spacings in the range $aT_c=1/2$ to
$1/8$ (for the Wilson action, $5.1<\beta \le 6.0$).
At large lattice spacing ($aT_c = 1/2$) the best signals came
from the Polyakov loop itself,
measured with the Parisi --
Petronzio -- Rapuano \cite{PPR} multihit variance reduction method.
At smaller lattice spacing ($aT_c \le 1/3$) the signal from this operator
degrades and we had more success using correlators of APE--blocked
\cite{APEBlock} links: we iterate
\begin{eqnarray}
V^{n+1}_j(x) = (1-\alpha)V^{n}_j(x) & +  & \alpha/4 \sum_{k \ne j}
(V^{n}_k(x)V^{n}_j(x+\hat k)V^{n}_j(x+\hat j)^\dagger
\nonumber  \\
& + & V^{n}_k(x- \hat k)^\dagger
 V^{n}_j(x- \hat k)V^{n}_j(x - \hat k +\hat j) )
\end{eqnarray}
(with $V^0_j(x)=U_j(x)$ and $V^{n+1}_j(x)$ projected back onto $SU(3)$), 
with $\alpha$ varying from 0.2 at $aT_c=1/3$ to
0.5 at $aT_c = 1/4$ to $0.7$ at $aT_c=1/8$
and the number of blocking steps rising from 10 to 15. Presumably a large
part of the noise is local (in physical units) and hence more nonlocal
(in units of $a$) averaging is needed as the lattice spacing falls.

The multihit algorithm is not very convenient when using the FP action
on a parallel machine for the same reason that Metropolis or 
overrelaxation
is inconvenient, but as we only ran the small lattice spacing simulations
on parallel machines that is not a problem.

We determined the torelon mass from a single-exponential correlated fit,
beginning at a minimum $z$ where the $\chi^2/DF$ is near unity, and where we
see stability in the effective mass over a range of $z$. At $aT_c = 1/2$
the fits are to $z=1$ and $z=2$ points only since there is no signal for $z>2$.

As a check of our programs we reproduced the torelon mass measurements of
Michael and Teper \cite{MT}
on Wilson $\beta=5.9$, $12^3$ and $\beta=6.0$, $16^3$
lattices. At lower $\beta$ there is little Wilson data for torelons which is
a variational bound to which we can compare: at $\beta=5.69$ our torelon
mass is greater than the $\beta=5.7$
 result of de Forcrand et.~al. \cite{DEFOR}
(which however was done using the non-variational cold wall source method.)




An uncertainty in the value of the critical coupling  propagates into
the uncertainty of $G$ for the FP action. 
By doing lower-statistics simulations at couplings
close to the critical $\beta$-values, we can roughly compute the
variation of the torelon mass with $\beta$. For the uncertainties in
$\beta_c$'s recorded in Table \ref{tab:betacrit}, this is an additional uncertainty
in $G$ of about 0.04--0.05 at
$r=2$ and $r=3/2$. We include this uncertainty in the figures
by combining it in quadrature with the statistical fluctuation in $G$;
in the figures the extreme range of the vertical error bars shows
the combined uncertainty.


\begin{table*}[hbt]
\setlength{\tabcolsep}{1.5pc}
\caption{Our measurements of torelon masses from the Wilson action on
small lattices.}
\label{tab:tor_w}
\begin{tabular*}{\textwidth}{@{}l@{\extracolsep{\fill}}lcccccc}
\hline
volume &  $\beta$ &  $\mu = L \sigma$ & $L\sqrt{\sigma}$\\
 \hline
 \hline
$3^3 \times 16$  & 5.10  & 1.83(3) & 2.34(2) \\
$4^3 \times 16$  & 5.10  & 2.86(6) & 3.38(4) \\
\hline
$6^3 \times 16$  & 5.55  & 1.65(5) & 3.15(5)  \\
$6^3 \times 16$  & 5.69  & 0.75(2) & 2.12(3) \\
\hline
$8^3 \times 16$  & 5.69  & 1.11(2) & 2.98(3) \\
$9^3 \times 16$  & 5.9   & 0.45(1) & 2.03(1) \\
\hline
$12^3 \times 16$  & 6.0  & 0.44(1) & 2.31(4) \\
\hline
\end{tabular*}
\end{table*}

\begin{table*}[hbt]
\setlength{\tabcolsep}{1.5pc}
\caption{Our measurements of torelon masses from the FP action.}
\label{tab:tor_afp}
\begin{tabular*}{\textwidth}{@{}l@{\extracolsep{\fill}}lcccccc}
\hline
volume &  $\beta$  & $\mu = L \sigma$ & $L\sqrt{\sigma}$\\
 \hline
 \hline
$3^3 \times 12$  & 2.975  & 1.50(6) & 2.12(4) \\
$4^3 \times 8$  & 2.975   & 2.38(5) & 3.09(3) \\
\hline
$6^3 \times 12$  & 3.34   & 1.55(5) & 3.05(5) \\
$6^3 \times 12$  & 3.52   & 0.78(1) & 2.16(1) \\
\hline
$8^3 \times 12$  & 3.52   & 1.19(4) & 3.08(4) \\
$9^3 \times 16$  & 3.79   & 0.50(1) & 2.12(2) \\
\hline
\end{tabular*}
\end{table*}

The result of our simulations is that the torelon mass measured
on aspect ratio 2 lattices using the FP action scales within the statistical
errors
for $1/4 \le aT_c \le 1/2$, at a value which is consistent with
the value inferred from
 Wilson action results from small lattice
spacing simulations.

As a second test of scaling we compare torelon masses measured on lattices
of constant aspect ratio 3/2. The scaling plot is shown in Fig.
\ref{fig:aspect_32a}.
In this figure the diamond shows the result of our simulation
with the Wilson action at $\beta=6.0$
and the star shows an extrapolation
to $\beta_c(N_t=8)=6.06$ using asymptotic freedom.
 All the simulations were performed by us.
Again, the  FP action torelon mass scales
for $1/6 \le aT_c \le 1/2$, while the Wilson action exhibits considerable
scaling violation over the same range of lattice spacings.

\subsection{The potential}

Finally, we have measured the potential from the correlation function of
Polyakov loops.The results obtained at $\beta_c(N_t=2)$ and 
$\beta_c(N_t=4)$ are shown in Figs. \ref{fig:wilson24} and 
\ref{fig:perfect24} for the Wilson and the 8 parameter FP actions,
respectively. At large distances the system should forget the errors 
related to the bad operators and rotational symmetry has to be
recovered. For the Wilson action at $N_t=2$ strong symmetry violation
is seen even at $rT_c=1.5$ and there is a clear scaling violation in
the long distance part of $V(r)$. These cut--off effects drop below
the statistical errors for the FP action.

\section{Conclusions}
We have described a systematic program for constructing fixed point
actions for SU(3) gauge theory, and illustrated it using a particular
scale two blocking transformation. An approximation to the FP action shows
scaling  within our small statistical errors 
beginning at $aT_c=1/2$, as compared to $aT_c \le 1/8$ for the
Wilson action, at a cost of a factor of about 7  in computational speed
per site update.

This work allows one to consider a large 
number of questions.

First, the action which is simulated here and shows scaling
 is an approximation to a FP action, not
to an action lying on any known renormalized trajectory
at small $\beta$. Can one find the renormalized trajectory of
these renormalization group transformations?
We believe that this is feasible using the available techniques
\cite{demon}.
As a related question, the FP action scales for $\beta$ values at
which it was not a good fit to the FP action. It would be useful to understand
how imperfect an action can be and still perform well.
Additionally, it might be that completely different
 scale transformations, such as the $\sqrt 3$ transformation \cite{SQRT3},
 might yield FP actions which are even more local than this one.
 
In the meantime, the fixed point action can be used for many
phenomenological calculations at large lattice spacing.

A particularly severe and unexpected bottleneck in this
calculation was the absence of good scaling tests using the Wilson action.
We are particularly hampered by imprecision in published values of
deconfinement couplings for $N_t \ge 8$.
We urge groups doing simulations with the Wilson action to choose parameters
which are easy to relate to other length scales.
We would also like to encourage developers of alternative improved
actions to test them at very large values of the lattice spacing, since either
scaling violations will be large and easy to see, or they will be small,
showing that the candidate action is indeed an improvement.
 
Finally we would like to comment on lattice perturbation theory.
 The construction of the FP  actions we have described here
 does not involve lattice perturbation theory in the sense that we have made no
expansions in $g^2$ or in powers of the lattice spacing $a$. The
actions themselves are not explicitly
designed to have good perturbative expansions,
although the one we have tested most extensively
 probably does so in terms of its
bare coupling (since it seems to scale asymptotically with a small
$T_c/\Lambda$ ratio). Perturbative calculations for these actions are
unpleasant but not impossible, given symbolic manipulation programs. 
 Direct numerical calculations of
renormalization constants \cite{SACHRAJDA}
 are in principle no more difficult than for
the Wilson action, although the currents themselves must be ``perfect.''
 
These actions appear to open the way towards doing QCD simulations
with  modest computer resources.

\section{Acknowledgements}
U. Wiese participated in the early stages of this project.
We would like to thank  T.~Barker,
M.~Horanyi and the Colorado high energy experimental
group for allowing us to use their work stations. Some computations were
performed on the Paragon at the San Diego Supercomputer Center
and on the T3D at NERSC, Livermore.
We want to thank P.~B\"uttiker, M.~Egger and M.~Willers for their 
support in using the computing
facilities at the University of Bern.
T. D. would like to thank C. DeTar and U. Heller for discussions about
measuring string tensions and D. Toussaint for help with the code.
This work was supported by the U.S. Department of 
Energy and by the National Science Foundation and by the Swiss National
Science Foundation.

\appendix

\section{Appendix}

We discuss here briefly the arguments which show that the numerical
analysis of the classical FP equation Eq.~(\ref{2}) can be done
on small lattices.

Consider a lattice of size $(mL)^4$ onto which a periodic $L^4$
coarse $V$ configuration is copied $m^4$ times, where $m$ is a large
integer. The corresponding minimizing configuration $U_0(V)$ lives
on an $(2mL)^4$ lattice. The fine links $U_{\mu}(n)$ and 
$U_{\mu}(n+2L \hat{\nu})$ are influenced identically by the coarse
field $V$. Assuming that this symmetry is not broken spontaneously,
the minimizing configuration will be periodic with period $2L$. The
relation between the coarse field and the minimizing fine field on
the big lattices is fixed by the relation between the $L^4$ and $(2L)^4$
lattice. On the other hand, the big $(mL)^4$ configurations are
certainly legitimate candidates to be used to parametrize the FP
action, since  Eq.~(\ref{2}) should be satisfied on any type of $V$ 
configurations.

The question remains whether the specific configurations considered above
give sufficient information to fix the coefficients in Eq.~(\ref{4}).
In spin models it is easy to see that solving the FP equation on small
lattices one is not able to resolve certain couplings in the FP
action. Working on a $2^2$ lattice in the non--linear $\sigma$--model,
for example, one can not separately determine the two--spin couplings
$\rho(1,0),\rho(3,0),\rho(5,0),...$ -- only their sum enters. Using a
limited number of loops in a non--Abelian gauge theory on a small lattice
it is not clear whether such relations exist. We emphasize, however
that even if they exist, no error is caused by small lattices, only
information is lost. Since the general parametrization in terms of loops
is highly redundant and our specific parametrization is not at all
systematic, this is not a real problem.

\newcommand{\PL}[3]{{Phys. Lett.} {\bf #1} {(19#2)} #3}
\newcommand{\PR}[3]{{Phys. Rev.} {\bf #1} {(19#2)}  #3}
\newcommand{\NP}[3]{{Nucl. Phys.} {\bf #1} {(19#2)} #3}
\newcommand{\PRL}[3]{{Phys. Rev. Lett.} {\bf #1} {(19#2)} #3}
\newcommand{\PREPC}[3]{{Phys. Rep.} {\bf #1} {(19#2)}  #3}
\newcommand{\ZPHYS}[3]{{Z. Phys.} {\bf #1} {(19#2)} #3}
\newcommand{\ANN}[3]{{Ann. Phys. (N.Y.)} {\bf #1} {(19#2)} #3}
\newcommand{\HELV}[3]{{Helv. Phys. Acta} {\bf #1} {(19#2)} #3}
\newcommand{\NC}[3]{{Nuovo Cim.} {\bf #1} {(19#2)} #3}
\newcommand{\CMP}[3]{{Comm. Math. Phys.} {\bf #1} {(19#2)} #3}
\newcommand{\REVMP}[3]{{Rev. Mod. Phys.} {\bf #1} {(19#2)} #3}
\newcommand{\ADD}[3]{{\hspace{.1truecm}}{\bf #1} {(19#2)} #3}
\newcommand{\PA}[3] {{Physica} {\bf #1} {(19#2)} #3}
\newcommand{\JE}[3] {{JETP} {\bf #1} {(19#2)} #3}
\newcommand{\FS}[3] {{Nucl. Phys.} {\bf #1}{[FS#2]} {(19#2)} #3}

\end{document}